\documentclass[aps,prb,showpacs,twocolumn]{revtex4}
\usepackage{amsfonts}
\usepackage{amssymb}
\usepackage{amsmath}
\usepackage{graphicx}
\usepackage{epsfig}
\begin{document}

\title{Mosaic spin models with topological order}
\author{S. Yang$^{1}$}
\author{D. L. Zhou$^{2}$}
\author{C. P. Sun$^{1}$}
\affiliation{$^1$Institute of Theoretical Physics, Chinese Academy
of Sciences, Beijing
100080, China\\
$^2$Institute of Physics, Chinese Academy of Sciences, Beijing
100080, China}

\begin{abstract}
We study a class of two-dimensional spin models with the Kitaev-type
couplings in mosaic structure lattices to implement topological
orders. We show that they are exactly solvable by reducing them to
some free Majorana fermion models with gauge symmetries. The typical
case with a 4-8-8 close packing is investigated in detail to display
the quantum phases with Abelian and non-Abelian anyons. Its
topological properties characterized by Chern numbers are revealed
through the edge modes of its spectrum.
\end{abstract}
\pacs{75.10.Jm, 05.30.Pr, 71.10.Pm}
\maketitle

\emph{Introduction-} The phenomenon of emergence (such as a phase
transition) in a condensed matter system is usually understood
according to the Landau symmetry-breaking theory
(LSBT)\cite{wen-book}. There also exists a new kind
of order called \textquotedblleft topological order\textquotedblright \cite%
{Kitaev,wen-book,Wen,preskill9,pachos07} which cannot be described
in the frame of the LSBT (e.g., fractional quantum Hall effect). The
study of topological order in theoretical and experimental aspects
has been an active area of
research\cite%
{Kitaev,Wen,preskill9,pachos07,Sarma,pachos,Bombin,XT,CHD,Bonesteel,Oritz,Freedman,Read,YuYue}
. Since local perturbations hardly destroy the topological
properties, such topologically ordered states show exciting
potential to encode and process quantum information
robustly\cite{Kitaev}. Therefore it is significant and challenging
to find more exactly solvable models showing various topological
orders.

In this Rapid Communication, the Kitaev's honeycomb
model\cite{Kitaev} is generalized to the general mosaic spin models
with different two-dimensional Bravais lattices of complex unit
cells. Then we study the 4-8-8 case in detail to reveal the general
and special properties of mosaic spin models.

\begin{figure}[tbp]
\includegraphics[bb=70 465 560 780, width=8 cm, clip]{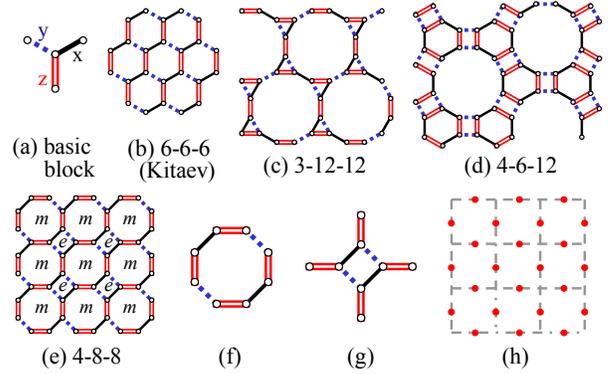}
\caption{(Color online) (a) Basic block for mosaic spin models,
which consists of three branches with x- (black solid link), y-
(blue dotted link), and z- (red double link) type couplings. (b)
6-6-6 mosaic model, i.e., Kitaev's honeycomb model. (c) 3-12-12
mosaic model. (d) 4-6-12 mosaic model. (e) 4-8-8 mosaic model, $e$
vortices lie on squares while $m$ vortices lie on octagons. (f) and
(g) The possible nonconstant terms of the effective Hamiltonian are
obtained by flipping four spin pairs around a octagon (f) and a
quatrefoil (g). (h) Kitaev's toric code model is the effective model
of the 4-8-8 mosaic model when $|J_{z}|\gg |J_{x}|,|J_{y}|$.}
\label{model}
\end{figure}

Our mosaic spin models are constructed with the basic block shown in Fig. %
\ref{model}(a), which is a vertex with three different types of spin
couplings along $x$- (black solid link), $y$- (blue dotted link),
and $z$- (red double link) directions, respectively. In spite of the
lattice symmetry, numerous spin models can be built based on this
basic block. However, taking translational symmetry and rotational
symmetry as much as possible into account, we regard each basic
block as the common vertex of three isogons with $n_{1}$, $n_{2}$
and $n_{3}$ edges, so there
are only four kinds of mosaic spin models\cite{TObook} illustrated in Fig. %
\ref{model}(b)-\ref{model}(e), called $n_{1}$-$n_{2}$-$n_{3}$ mosaic
models.

Obviously, the 6-6-6 mosaic model is just Kitaev's honeycomb model\cite%
{Kitaev}. Here, we remark that for given $n_{1}$, $n_{2}$, and
$n_{3}$, there exist some unequivalent kinds of plane arrangement of
$x$ links, $y$ links, and $z$ links, but we only illustrate one of
them in Fig. \ref{model}. The general Hamiltonian of all mosaic spin
models reads as
\begin{equation}
H=-\sum_{u=x,y,z}J_{u}\sum_{(j,k)\in S(u)}\sigma _{j}^{u}\sigma
_{k}^{u} \label{H},
\end{equation}
where $S(u)$ is the set of links with $u$-direction couplings.

\emph{Perturbation theory study and abelian anyons - }First, we
study the 4-8-8 mosaic model as a typical illustration in detail. To
see its topological properties, we first analyze its low energy
excitations when the system is initially spontaneously polarized
with the strong Ising interaction $H_{0}=-J_{z}\sum_{z links}\sigma
_{j}^{z}\sigma _{k}^{z}$. The ground energy of $H_{0}$ is
$E_{0}=-NJ_{z}$, where $N$ is the number of $z$ links. For larger
$J_{z}$ in comparison with $J_{x}$ and $J_{y}$, we regard the
transverse part $V=H-H_{0}$ as a perturbation and then prove that
the obtained effective Hamiltonian $H_{eff}$ just describes Kitaev's
toric code model\cite{Kitaev}, which supports many topological
issues of the original mosaic spin model.

The ground eigenstates of $H_{0}$ are highly degenerate, where each
two spins connected by a $z$ link can be either $\left\vert \uparrow
\uparrow \right\rangle $ or $\left\vert \downarrow \downarrow
\right\rangle $. The fusion projection\cite{Kitaev} $\Upsilon
_{l}^{\dag }$ can map the $l$th
aligned spin pair $\left\vert m,m\right\rangle _{l}$ to an effective spin $%
\left\vert m\right\rangle _{l}$ ($m=\uparrow $ or $\downarrow $), i.e., $%
\Upsilon _{l}^{\dag }\left\vert m,m\right\rangle _{l}=\left\vert
m\right\rangle _{l}$. We use the fusion projection and the Green function
formalism to calculate the effective Hamiltonian $H_{eff}=\sum_{l=0}^{\infty
}H_{eff}^{\left( l\right) }$ $=E_{0}+\Upsilon ^{\dag }V[1+G_{0}\left(
E_{0}\right) +G_{0}\left( E_{0}\right) VG_{0}\left( E_{0}\right) ]V\Upsilon
+\cdots $ where $G_{0}\left( E_{0}\right) =\left( E_{0}-H_{0}\right) ^{-1}$.
We first obtain the constant zeroth order one, the vanishing first order and
third order\ ones. Here, each terms $\sigma _{j}^{x}\sigma _{k}^{x}$ or $%
\sigma _{j}^{y}\sigma _{k}^{y}$ in $V$ flips two spins, increasing the
energy by $4J_{z}$. Up to the second order perturbation, one $V$ flips two
spins and the other $V$ flips them back, giving $%
H_{eff}^{(2)}=-N(J_{x}^{2}+J_{y}^{2})/(4J_{z})$ as a constant. As
shown in Figs. \ref{model}(f) and \ref{model}(g), we take two
$\sigma _{j}^{x}\sigma _{k}^{x}$ and two $\sigma _{j}^{y}\sigma
_{k}^{y}$ from four $V$ around one octagon\ or one quatrefoil in a
particular order. Taking all the $2\times 4!=48$ possible cases into
account, we obtain the fourth order effective Hamiltonian
\begin{equation}
H_{eff}^{\left( 4\right) }=-\frac{J_{x}^{2}J_{y}^{2}}{16J_{z}^{3}}
(5\sum_{O}\sigma _{l}^{y}\sigma _{r}^{y}\sigma _{u}^{y}\sigma
_{d}^{y}+\sum_{Q}\sigma _{l}^{z}\sigma _{r}^{z}\sigma _{u}^{z}\sigma
_{d}^{z})  \label{Heff},
\end{equation}%
where the constant term was dropped, $O$ and $Q$ represent the
octagon and quatrefoil in the two-dimensional (2D) lattice. Up to a
unitary transformation for spin rotation $\sigma ^{y}\rightarrow
\sigma ^{z}$, $\sigma ^{z}\rightarrow \sigma ^{x}$, $\sigma
^{x}\rightarrow \sigma ^{y}$, the above Hamiltonian represents the
Kitaev's toric code model\cite{Kitaev}. Thus the above
fusion projection constructs a new Bravais lattice illustrated in Fig. \ref%
{model}(h) with the effective spins laying on its links. Considering
Kitaev model (\ref{Heff}) possesses rich topological features
characterized by $m$ and $e$ anyons, we conclude that $m$ particles
live on octagons while $e$ particles live on squares in our model
with original spin representation.

\begin{figure}[tbp]
\includegraphics[bb=52 460 555 760, width=8 cm, clip]{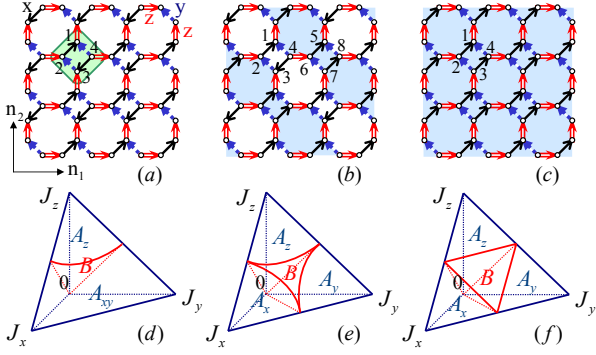}
\caption{(Color online) (a)-(c) 4-8-8 mosaic spin models in (a)
vortex-free (VF) sector, (b) vortex-half occupied (VHO) sector, and
(c) vortex-full occupied (VFO) sector. (d)-(f) The corresponding
phase graphs of the above lattices with gapless phase $B$ and gapped
phases $A$: (d) VF, (e) VHO, and (f) VFO.} \label{vortex}
\end{figure}

\emph{Majorana fermion mapping with }$%
\mathbb{Z}
_{2}$\emph{-gauge symmetry - } The 4-8-8 mosaic model consists of
four equivalent simple sublattices, and a unit cell [see the green
rhombus tablet in Fig. \ref{vortex}(a)] contains each of four kinds
of vortices referred to as 1, 2, 3, and 4. According to
Kitaev\cite{Kitaev}, we use the Majorana fermion operators to
represent Pauli operators as $\sigma ^{x}=ib^{x}c$, $\sigma
^{y}=ib^{y}c$, and $\sigma ^{z}=ib^{z}c$, where Majorana operators $%
b^{x},b^{y},b^{z}$, and $c$ satisfy $\alpha ^{2}=1$, $\alpha \beta
=-\beta
\alpha $ for $\alpha ,\beta \in \left\{ b^{x},b^{y},b^{z},c\right\} $ and $%
\alpha \neq \beta $. Then, the Hamiltonian (\ref{H}) can be rewritten as $%
H=\sum_{j,k}\frac{1}{2}G_{jk}c_{j}c_{k}$, where the operator-valued coupling
$G_{jk}\equiv iJ_{u}Z_{jk}$ ($u=x,y,z$) if $(j,k)\in S(u)$; $G_{jk}=0$ when $%
(j,k)\notin S(u)$. Here, a link $(j,k)$ determines a type of coupling $%
u=u(j,k)$. Due to the vanishing anticommutator of $b_{j}^{u}$ and $b_{k}^{u}$
, we have $Z_{jk}=-Z_{kj}$ for $j\neq k$.

For each site, the above-mentioned Majorana operators act on a $4D$
space, but the physical subspace is only $2D$. Thus we need to
invoke a gauge
transformation of $%
\mathbb{Z}
_{2}$ group to project the extended space into the physical subspace through
the projection operator$\ D=b^{x}b^{y}b^{z}c$: $\left\vert \psi
\right\rangle $ belongs to the physical subspace if and only if $D\left\vert
\psi \right\rangle =\left\vert \psi \right\rangle $. With this physical
projection, some eigenstates of $H$ can be found exactly because $G_{jk}$
lays on the center of an Abelian algebra generated by $Z_{jk}$ with $\left[
Z_{jk},H\right] =0$ and $\left[ Z_{jk},Z_{ml}\right] =0$. Since $%
(Z_{jk})^{2}=1,Z_{jk}=ib_{j}^{u}b_{k}^{u}$ generates a $%
\mathbb{Z}
_{2}$ group and its eigenvalues are $z_{jk}=\pm 1.$ Therefore,
$\left\{
Z_{jk},I|(j,k)\in S(u),u=x,y,z\right\} $ generate the symmetry group $%
\mathbb{Z}_{2}\otimes \mathbb{Z}_{2}\otimes \cdots \otimes
\mathbb{Z}_{2}$ of the model; the whole Hilbert space is then
decomposed according to the direct sum of some irreducible
representations, and each irreducible sector is characterized by
$\{z_{jk}|(j,k)\in S(u),u=x,y,z\}$, i.e., the directions shown in
Figs. \ref{vortex}(a)-\ref{vortex}(c).

Obviously, in each irreducible representation space, we can reduce the
Hamiltonian (\ref{H}) into a quadratic form, which represents an effective
Hamiltonian of free fermions for a given vortex arrangement. To characterize
the vortex configuration, we introduce square and octagon plaquette
operators $W_{p}^{\left( 4\right) }=\sigma _{1}^{z}\sigma _{2}^{z}\sigma
_{3}^{z}\sigma _{4}^{z}$ and $W_{p}^{\left( 8\right) }=\sigma _{1}^{y}\sigma
_{2}^{y}\sigma _{3}^{x}\sigma _{4}^{x}\sigma _{5}^{y}\sigma _{6}^{y}\sigma
_{7}^{x}\sigma _{8}^{x}$ or
\begin{equation}
W_{p}^{\left( 4\right) }=-\prod_{\left( j,k\right) \in \partial
p(4)}Z_{jk}, W_{p}^{\left( 8\right) }=-\prod_{\left( j,k\right) \in
\partial p(8)}Z_{jk},
\end{equation}
where $\partial p(4)$ and $\partial p(8)$ represent the sets of
boundary links of square and octagon\ plaquettes with label $p$; the
$\left( j,k\right) $ links are ordered clockwise around the
plaquette. The operators $W_{p}^{\left( j\right) }$ ($j=4,8$)
commute with each other, $\left[ W_{p}^{\left( j\right) },H\right]
=0$, $W_{p}^{\left( 4\right) 2}=W_{p}^{\left( 8\right) 2}=I$, and
thus each plaquette operator has two eigenvalues $w_{p}=\pm 1$. A
plaquette with $w_{p}=1$ is a vortex-free plaquette while $w_{p}=-1$
corresponds to a vortex. In the following we will show that
different arrangements of vortices result in different phase graphs
and different energy spectrums.

\emph{\ 4-8-8 mosaic model in different vortex-occupied sectors - }Let us
denote the site index $j$ in detail by $\left( s,\lambda \right) $, where $%
s $ refers to a unit cell, and $\lambda $ to a position type inside the
cell. The Hamiltonian then reads $H=\sum_{s,\lambda ,t,\mu }G_{s\lambda
,t\mu }c_{s\lambda }c_{t\mu }/2$. Due to the translational invariance of the
lattice along the unit direction vectors $\mathbf{n}_{1}=\left( 1,0\right) $
, $\mathbf{n}_{2}=\left( 0,1\right) $, $G_{s\lambda ,t\mu }$ actually
depends on $\lambda ,\mu $ and $t-s$, and thus $\exp \left[ i\mathbf{q}\cdot
\left( \mathbf{r}_{t}-\mathbf{r}_{s}\right) \right] G_{s\lambda ,t\mu }=\exp
\left( i\mathbf{q}\cdot \mathbf{r}_{t}\right) G_{0\lambda ,t\mu }$. To study
the spectral structure of the system, we invoke the generic fermion operator
$a_{\mathbf{q},\mu }$ $=\sum_{t}e^{i\mathbf{q}\cdot \mathbf{r}_{t}}c_{t\mu
}/ \sqrt{2N}$ where $N$ is the total number of the unit cells and $a_{%
\mathbf{p} ,\mu }a_{\mathbf{q},\lambda }^{\dag }+a_{\mathbf{q},\lambda
}^{\dag }a_{ \mathbf{p},\mu }=\delta _{\mathbf{pq}}\delta _{\mu \lambda }$. $%
\widetilde{G} _{\lambda \mu }\left( \mathbf{q}\right) $ is the
Fourier transformation of $G_{0\lambda ,t\mu }$. In the momentum
space, the fermion representation of the Hamiltonian reads
\begin{equation}
H=\frac{1}{2}\sum_{\mathbf{q}}A_{\mathbf{q}}^{\dag
}\widetilde{G}\left( \mathbf{q}\right) A_{\mathbf{q}}.
\end{equation}

Case I: In the vortex-free (VF) sector, we choose a particular direction ($%
z_{jk}=+1$ or $-1$) for each link [see Fig. \ref{vortex}(a)], so
that translational symmetry holds and $w_{p}^{\left( 4\right)
}=w_{p}^{\left(
8\right) }=1$ for all plaquettes. Since $A_{\mathbf{q}}^{\dag }=(a_{\mathbf{%
q },1}^{\dag },a_{\mathbf{q},2}^{\dag },a_{\mathbf{q},3}^{\dag },a_{\mathbf{q%
} ,4}^{\dag })$, we have the $4\times 4$ spectral matrix $\widetilde{G}%
\left( \mathbf{q}\right) =\widetilde{G}_{VF}$ or
\begin{equation}
\widetilde{G}_{VF}=\left(
\begin{array}{cc}
J_{x}\sigma ^{y} & -iJ_{y}\sigma ^{x}+iJ_{z}\alpha \\
iJ_{y}\sigma ^{x}-iJ_{z}\alpha ^{\dag } & J_{x}\sigma ^{y}%
\end{array}
\right),
\end{equation}
where $\alpha =diag.[\exp (-iq_{2}),-\exp (iq_{1})]$,
$q_{1}=\mathbf{q\cdot n }_{1}$, $q_{2}=\mathbf{q\cdot n}_{2}$.

\begin{figure}[tbp]
\includegraphics[bb=53 590 540 770, width=8 cm, clip]{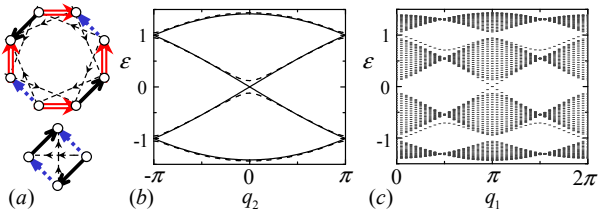}
\caption{(Color online) (a) Thin dashed arrows describe the
effective second nearest-neighbor interactions between fermions and
the corresponding gauge
induced by a magnetic field. (b) Profile graph of an energy spectrum with $%
J_{x}=J_{y}=1$, $J_{z}=\protect\sqrt{2}$ along the
$q_{1}=\protect\pi $ axis in the absence (solid lines) and presence
(dashed lines) of a magnetic field. (c) Energy spectrum of the above
system with finite size along the $\mathbf{n}_{2}$ direction in the
magnetic field. Two chiral edge modes crossing at $E=0$ correspond
to Chern number $\pm 1$.} \label{free}
\end{figure}

The single particle spectrum $\varepsilon \left( \mathbf{q}\right)
=-\varepsilon \left( -\mathbf{q}\right) $ is given by the
eigenvalues of the spectral matrix $\widetilde{G}\left(
\mathbf{q}\right) $. An important property of the spectrum is
whether it is gapless, i.e., whether $ \varepsilon \left(
\mathbf{q}\right) $ vanishes for some $\mathbf{q}$. Obviously, the
vanishing of determinant $Det(\widetilde{G}_{VF})$ enjoys the zero
eigenvalues of $\widetilde{G}_{VF}$. Then the gapless condition is
\begin{equation}
J_{x}^{2}+J_{y}^{2}=J_{z}^{2}.
\end{equation}
As shown in Fig. \ref{vortex}(d), the phase diagram of our model consists of
three phases, the gapless phase $B$, which is actually a conical surface,
distinguishing from two gapped phases $A_{z}$ and $A_{xy}$. Since the
possible zero energy degenerate points are $\left( 0,\pm \pi \right) $ and $%
\left( \pm \pi ,0\right) $ in the first Brillouin zone, we choose $%
J_{x}=J_{y}=1$, $J_{z}=\sqrt{2}$, and $q_{1}=\pi $ to plot the
profile graph of the energy spectrum with respect to $q_{2}\in
\left[ -\pi ,\pi \right] $ in Fig. \ref{free}(b) by solid lines. The
eigenvalues of $\widetilde{G}_{VF}$
are chosen in the concourse \{$\pm \sqrt{2}\cos \left( q_{2}/4\right) $, $%
\pm \sqrt{2}\sin \left( q_{2}/4\right) $\}. Thus in the vicinity of
the energy degenerate points, the low-energy excited spectrum is
approximately linear. This property maybe helpful to study quantum
state transfer problems\cite{YS13}.

Case II: We choose another particular direction for each link as
shown in Fig. \ref{vortex}(b), and the plaquettes with
$w_{p}^{\left( 4\right) }=-1$ or $w_{p}^{\left( 8\right) }=-1$ are
marked by blue shadings. In this vortex-half occupied (VHO) lattice,
each unit cell contains eight kinds of sites, $A_{\mathbf{q}}^{\dag
\prime }=(a_{\mathbf{q},1}^{\dag },a_{\mathbf{q}
,2}^{\dag },a_{\mathbf{q},3}^{\dag },a_{\mathbf{q},4}^{\dag }$, $a_{\mathbf{%
q },5}^{\dag },a_{\mathbf{q},6}^{\dag },a_{\mathbf{q},7}^{\dag },a_{\mathbf{q%
} ,8}^{\dag })$, the corresponding $8\times 8$ spectral matrix becomes
\begin{equation}
\widetilde{G}_{VHO}=\left(
\begin{array}{cccc}
J_{x}\sigma _{y} & -iJ_{y}\sigma _{x} & 0 & iJ_{z}e^{-iq_{2}^{\prime
}}\alpha ^{\dag } \\
iJ_{y}\sigma _{x} & J_{x}\sigma _{y} & -iJ_{z}\beta ^{\dag } & 0 \\
0 & iJ_{z}\beta & J_{x}\sigma _{y} & -iJ_{y}\sigma _{x} \\
-iJ_{z}e^{iq_{2}^{\prime }}\alpha & 0 & iJ_{y}\sigma _{x} & -J_{x}\sigma _{y}%
\end{array}
\right),
\end{equation}
where $\alpha =diag\left( 1,-e^{-iq_{1}^{\prime }}\right) $, $\beta
=diag\left( e^{-iq_{1}^{\prime }},-1\right) $, $q_{1}^{\prime
}=\mathbf{\ q\cdot n}_{1}^{\prime }$, $q_{2}^{\prime
}=\mathbf{q\cdot n}_{2}^{\prime }$, $\mathbf{n}_{1}^{\prime }=\left(
1,1\right) $, and $\mathbf{n}_{2}^{\prime }=\left( -1,1\right) $.
The gapless condition for VHO lattice is
\begin{equation}
J_{x}^{2}<J_{y}^{2}+J_{z}^{2},J_{y}^{2}<J_{x}^{2}+J_{z}^{2},J_{z}^{2}<J_{x}^{2}+J_{y}^{2}
\end{equation}
and the corresponding phase graph is plotted in Fig.
\ref{vortex}(e). We notice that the same phase graph has been
obtained by Pachos\cite{pachos} for the Kitaev model.

Case III: We choose the directions of links as shown in Fig.
\ref{vortex}(c) so that the translational symmetry still holds and
$w_{p}^{\left( 4\right) }=w_{p}^{\left( 8\right) }=-1$ for every
plaquette. The unit cell can be chosen as the same as the one in the
VF sector, so do $\alpha $, $q_{1}$, and $q_{2}$. Therefore
\begin{equation}
\widetilde{G}_{VFO}=\left(
\begin{array}{cc}
J_{x}\sigma ^{y} & -iJ_{y}\sigma ^{x}+iJ_{z}\alpha \\
iJ_{y}\sigma ^{x}-iJ_{z}\alpha ^{\dag } & -J_{x}\sigma ^{y}%
\end{array}
\right).
\end{equation}
The gapless condition is found as
\begin{equation}
\left( J_{x}-J_{z}\right) ^{2}\leq J_{y}^{2}\leq \left( J_{x}+J_{z}\right)
^{2}.
\end{equation}
If $J_{x},J_{y},J_{z}\geq 0$, we have $J_{x}\leq J_{y}+J_{z}$,
$J_{y}\leq J_{x}+J_{z}$, $J_{z}\leq J_{x}+J_{y}$. Thus in this case
the phase diagram of our model is the same as that of Kitaev's
honeycomb model. As shown in Fig. \ref{vortex}(f), the region within
the red lines labeled by $B$ is gapless. The other three gapped
phases $A_{x}$, $A_{y}$, and $A_{z}$ are algebraically distinct.
However, the energy spectrum of the 4-8-8 mosaic model is more
complex than that of the Kitaev model. When $J_{x}=J_{y}=J_{z}=1$
and $q_{1}=-q_{2}=q$, the eigenvalues of the single fermion are
chosen in the concourse \{$-1/2\pm \cos (q/2+\pi /4)$, $1/2\pm \cos
(q/2-\pi /4)$\}. Similarly, the different energy spectrums of mosaic
spin models imply their different dynamic properties.

In order to see the stability of the ground state in the VF sector,
we compare the ground energy $E_{0}=-\sum_{\mathbf{q}}\varepsilon
_{\mathbf{q}}/2$ of the VF lattice with that of the VHO and VFO
lattice mentioned above. By
choosing $J_{x}=J_{y}=J_{z}=1$, we find the ground energy per site is $%
E_{0,VF}=-0.80415$, $E_{0,VHO}=-0.75930$, and $E_{0,VFO}=-0.73631$, so $%
E_{0,VF}<E_{0,VHO}<E_{0,VFO}$. The other cases can be studied similarly.
Actually, as it was pointed out by Kitaev\cite{Kitaev}, Lieb's theorem\cite%
{Lieb} ensures that the VF lattice has the lowest energy to form a
ground state. In the following, we will focus on the stable VF
lattice and investigate the nontrivial topological properties in the
$B$ phase.

\emph{Topological properties of B phase in the presence of magnetic
field - } The perturbation $V=-\sum_{j}\left( h_{x}\sigma
_{j}^{x}+h_{y}\sigma _{j}^{y}+h_{z}\sigma _{j}^{z}\right) $
introduced by Kitaev\cite{Kitaev} can break the time-reversal
symmetry. Then the nontrivial third-order term becomes
$H_{eff}^{\left( 3\right) }$ $=\kappa \sum_{j,k,l}$ $\left(
iZ_{jl}Z_{kl}\right)$ $ c_{j}c_{k}$, where $\kappa \sim
h_{x}h_{y}h_{z}/J^{2}$. As illustrated in Fig. \ref{free}(a), the
thin dashed arrows represent the effective second nearest-neighbor
interactions between fermions induced by $H_{eff}^{\left( 3\right)
}$, and their directions denote the chosen gauge $Z_{jl}Z_{kl}$.
When $\kappa =0.025$ , the changed profiled spectrum is figured by
dashed lines in Fig. \ref{free}(b). Therefore the system in the $B$
phase acquires an energy gap in the presence of a magnetic field,
which is helpful for protecting non-Abelian anyons.

According to Kitaev\cite{Kitaev}, the topological properties of a
two-dimensional noninteracting \ fermion system with an energy gap
are usually characterized by Chern number, which can be determined
by observing the edge modes of the spectrum\cite{Kitaev,Hatsugai}.
If the system illustrated in Fig. \ref{vortex}(a) is finite along
the $\mathbf{n}_{2}$ direction while still periodic in the
$\mathbf{n}_{1}$ direction, its energy spectrum is shown in Fig.
\ref{free}(c) with $J_{x}=J_{y}=1$, $J_{z}=\sqrt{2}$, and $\kappa
=0.025$. Then we can observe two edge modes (corresponding to two
edges) crossing at $E=0$. Since the Fermi energy lies in the central
gap, only these two edge states around zero energy are relevant to
Chern number\cite{Hatsugai}. We also notice that the two edge modes
have a universal chiral feature\cite{Kitaev}, i.e., even if the
edges are changed, the energy curves of the edge modes do not change
their tendencies, respectively.
Therefore we conclude that the Chern number is $\pm 1$ in the non-Abelian $%
B $ phase. We also get zero Chern number in the Abelian phases $A_{xy}$ and $%
A_{z}$ with similar studies. Compared with Kitaev's honeycomb model
and the 4-8-8 mosaic model with even cycles in the lattice, the
3-12-12 mosaic model with odd cycles spontaneously breaks time
reversal symmetry to obtain Chern number $\pm 1$ without applying a
magnetic field \cite{Kitaev,YaoHong}.

\emph{Conclusion - }We generalize Kitaev's honeycomb model to
various mosaic spin models with translation and rotation symmetries
and study the 4-8-8 case in detail. It is found that when
$|J_{z}|\gg|J_{x}|,|J_{y}|$, our model is equivalent to Kitaev's
toric code model with Abelian anyons. Different vortex excitations
result in different phase diagrams with a gapless and gapped
spectral structure. In the stable vortex-free case, the zero-energy
Dirac points appear and the external magnetic field can induce an
energy gap. The nontrivial Chern number in $B$ phase is obtained by
studying the edge modes of the spectrum.

We thank X. G. Wen, T. Xiang, Y. S. Wu, Y. Yu, H. Q. Lin, G. M.
Zhang, and J. Vidal for helpful discussions. One of the authors
(D.L.Z.) acknowledges the hospitality of the Program on
\textquotedblleft Quantum Phases of Matter\textquotedblright\ by
KITPC. The project was supported by the NSFC (Grants No. 90203018,
No. 10474104, and No. 60433050) and the NFRPC (Grants No.
2006CB921206, No. 2005CB724508, and No. 2006AA06Z104).

When this work is nearly finished, we notice that H. Yao and S. A.
Kivelson have just studied the 3-12-12 mosaic model in detail
\cite{YaoHong}.


\begin{thebibliography}{99}
\bibitem{wen-book} X. G. Wen, \textit{Quantum Field Theory of Many-Body
Systems }(Oxford University, New York, 2004).
\bibitem{Kitaev} A. Kitaev, Ann. Phys. (N.Y.) \textbf{303}, 2 (2003); \textbf{321}, 2 (2006).
\bibitem{Wen} X. G. Wen, Phys. Rev. Lett. \textbf{90}, 016803 (2003); M. A.
Levin and X. G. Wen, Phys. Rev. B \textbf{71}, 045110 (2005).
\bibitem{preskill9} J. Preskill, \textit{Topological quantum computation},
http:// www.theory.caltech.edu/people/preskill/ph229/ (2004).
\bibitem{pachos07} G. K. Brennen and J. K. Pachos, arXiv: 0704.2241,
Proc. R. Soc. London, Ser. A (to be published).
\bibitem{Sarma} S. D. Sarma, M. Freedman, C. Nayak, S. H. Simon, A. Stern,
arXiv: 0707.1889.
\bibitem{pachos} J. K. Pachos, Int. J. Quantum Inf. \textbf{4}, 947 (2006); Ann.
Phys. (N.Y.) \textbf{322}, 1254 (2007).
\bibitem{Bombin} H. Bombin and M. A. Martin-Delgado, Phys. Rev. Lett.
\textbf{97}, 180501 (2006).
\bibitem{XT} X. Y. Feng, G. M. Zhang, and T. Xiang, Phys. Rev. Lett. \textbf{%
\ 98}, 087204 (2007); D. H. Lee, G. M. Zhang, and T. Xiang, arXiv:
0705.3499, Phys. Rev. Lett. (to be published).
\bibitem{CHD} H. D. Chen and J. P. Hu, arXiv: cond-mat/0702366; H. D. Chen
and Z. Nussinov, arXiv: cond-mat/0703633.
\bibitem{Bonesteel} N. E. Bonesteel, L. Hormozi, G. Zikos, and S. H. Simon,
Phys. Rev. Lett. \textbf{95}, 140503 (2005); L. Hormozi, G. Zikos,
N. E. Bonesteel, and S. H. Simon, Phys. Rev. B \textbf{75}, 165310
(2007).
\bibitem{Oritz} Z. Nussinov, G. Ortiz, arXiv: cond-mat/0702377.
\bibitem{Freedman} L. Fidkowski, M. Freedman, C. Nayak, K. Walker, and Z. H.
Wang, arXiv: cond-mat/0610583.
\bibitem{Read} N. Read and D. Green, Phys. Rev. B \textbf{61}, 10267 (2000).
\bibitem{YuYue} Y. Yu, Z. Q. Wang, arXiv: 0708.0631.
\bibitem{TObook} R. Kerner, Chapter 3 of \textit{Topology in Condensed
Matter } (Springer, Berlin, 2006).
\bibitem{YS13} S. Yang, Z. Song, and C. P. Sun, Phys. Rev. A \textbf{73},
022317 (2006); Eur. Phys. J. B \textbf{52}, 377 (2006).
\bibitem{Lieb} E. H. Lieb, Phys. Rev. Lett. \textbf{73}, 2158 (1994).
\bibitem{Hatsugai} Y. Hatsugai, Phys. Rev. Lett. \textbf{71}, 3697 (1993);
Phys. Rev. B \textbf{48}, 11851 (1993).
\bibitem{YaoHong} H. Yao, S. A. Kivelson, arXiv: 0708.0040, Phys. Rev. Lett. (to be published).
\end{thebibliography}
\end{document}